\documentclass[aps,prl,twocolumn,showpacs]{revtex4}
\usepackage{epsfig}
\usepackage{times}
\begin{document}

\title{Defense mechanisms of empathetic players in the spatial ultimatum game}

\author{Attila Szolnoki,$^1$ Matja{\v z} Perc,$^2$ Gy{\"o}rgy Szab{\'o}$^1$}
\affiliation{$^1$Institute of Technical Physics and Materials Science, Research Centre for Natural Sciences, Hungarian Academy of Sciences, P.O. Box 49, H-1525 Budapest, Hungary \\
$^2$Faculty of Natural Sciences and Mathematics, University of Maribor, Koro{\v s}ka cesta 160, SI-2000 Maribor, Slovenia}

\begin{abstract}
Experiments on the ultimatum game have revealed that humans are remarkably fond of fair play. When asked to share an amount of money, unfair offers are rare and their acceptance rate small. While empathy and spatiality may lead to the evolution of fairness, thus far considered continuous strategies have precluded the observation of solutions that would be driven by pattern formation. Here we introduce a spatial ultimatum game with discrete strategies, and we show that this simple alteration opens the gate to fascinatingly rich dynamical behavior. Besides mixed stationary states, we report the occurrence of traveling waves and cyclic dominance, where one strategy in the cycle can be an alliance of two strategies. The highly webbed phase diagram, entailing continuous and discontinuous phase transitions, reveals hidden complexity in the pursuit of human fair play.
\end{abstract}

\maketitle

Imagine two players having to share a sum of money. One proposes a split, and the other can either agree with it or not. No haggling is allowed. If there is an agreement, the sum is shared according to the proposal. If not, both players remain empty handed. This is the blueprint of the ultimatum game, as proposed by G{\"u}th \textit{et al.} \cite{guth_jebo82}. Therein, a rational proposer should always claim the large majority of the sum, given that the responder ought to accept even the smallest amount offered. Experiments, however, reveal a different reality; one where the selfish and fully rational \textit{Homo economicus} frequently gives way to the emotional \textit{Homo emoticus} \cite{sigmund_sa02}. In fact, largely regardless of sex, age and the amount of money at stake, people refuse to accept offers they perceive as too small \cite{thaler_jep88, bolton_geb95}. More precisely, offers below one third of the total amount are rejected as often as they are accepted, and not surprisingly, more than two thirds of all offers will be remarkably close to the fair 50:50 split.

While explanations for the human fondness of fair division range from the psychologically-inspired definitions of utility functions \cite{kirchsteiger_jebo94} to the failure of ``seizing the moment'' (implying that after the game there will be no further interactions between the two players) \cite{fehr_qje99}, theoretical studies indicate that empathy \cite{page_bmb02, sanchez_jtb05}, spatial structure \cite{page_prsb00, killingback_prsb01, iranzo_jtb11}, heterogeneity \cite{da-silva_r_bjp07}, and reputation \cite{nowak_s00} play a pivotal role. In particular, Page \textit{et al.} \cite{page_prsb00} have shown that in well-mixed populations natural selection favors the rational solution, while spatiality may lead to much fairer outcomes. This result has been tested thoroughly against different types of players and updating rules \cite{iranzo_jtb11}, on various interaction networks \cite{kuperman_epjb08, eguiluz_acs09, li_x_pre09, sinatra_jstat09, xianyu_b_pa10b}, as well as under coevolution \cite{gao_j_epl11}. Although there is no doubt that spatial structure promotes the evolution of fairness, it is interesting that even in a non-spatial setting fair play may evolve if the population is small \cite{huck_geb99} or if players are empathetic, \textit{i.e.} if their offer $p$ matches the acceptance level $q$ \cite{page_bmb02}.

\begin{figure}[ht]
\begin{center} \includegraphics[width = 6.4cm]{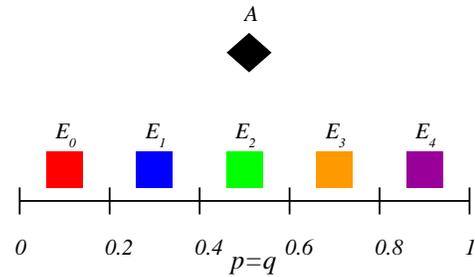}
\caption{\label{scheme}Schematic presentation of five discrete empathetic strategies $E_i$ ($i=0, 1, 2, 3, 4$) with the possible values of $p=q$ spread across the unit interval. The additional strategy $A$ is characterized by an arbitrary $(p,q)$ pair. While in the absence of $A$ players adopting the fair $E_2$ strategy are the undisputed victors, the outcome changes dramatically when the $A$ strategy is present, and it depends intricately on the $(p,q)$ pair that characterizes it.}
\end{center}
\end{figure}

Apart from occasionally studied minigame versions of the ultimatum game \cite{nowak_s00}, which consider only a few representative strategies and are traditionally employed to demonstrate principles of general importance, previous works focused on the full continuum of strategies as given by $p,q \in [0,1]$ (without loss of generality assuming that the sum to be divided equals one). Here we depart from this concept by proposing a spatial ultimatum game with discrete strategies in order to reveal solutions that are driven by pattern formation. Unlike minigames, which typically feature a small number of carefully chosen strategies, we still consider the whole unit interval of $p$ and $q$. But instead of the infinite number of continuous strategies, we introduce $N$ discrete strategies $E_i$ where $i=0, 1, \ldots, N-1$, as schematically depicted in Fig.~\ref{scheme} for $N=5$. We consider $E_i$ to be empathetic players and thus characterized by $p=q$, yet contested by an additional strategy $A$ that is characterized by an arbitrary $(p,q)$ pair. Initially, we thus have $N+1$ strategies that are distributed at random with equal probability on a $L^2$ square lattice, whereby each player $x$ is assigned a $(p,q)$ pair, corresponding to its offering and acceptance level. For players adopting strategy $A$ we will consider the fixed $p$ and $q$ values as the two main parameters determining the evolutionary outcome of the game, while a player $x$ adopting one of the $E_i$ ($i=0,1, \ldots, N-1$) empathetic strategies has $p_x=q_x=(r+i)/N$ where the random real number $r \in [0,1)$ is drawn for the creation of the strategy (either in the initial state or during the strategy adoption process).

The evolution of the initial strategy distribution is performed by repeating the following elementary steps in accordance with the Monte Carlo simulation procedure. First, a randomly selected player $x$ acquires its payoff $U_x$ by playing the ultimatum game with its four nearest neighbors, whereby in each pairwise interaction acting once as proposer with $p_x$ and once as a responder with $q_x$. Next, a randomly chosen neighbor, denoted by $y$, also acquires its payoff $U_y$ in the same way. Lastly, player $x$ tries to enforce its strategy on player $y$ in accordance with the probability $w=\{1+\exp[(U_{y}-U_{x})/K]\}^{-1}$, where $K$ quantifies the amplitude of noise \cite{szabo_pr07}. Without loosing generality, we set $K=0.1$, making it very likely that better performing players will pass their strategy onto their neighbors, yet it is also possible that players will occasionally learn from those performing worse.

We emphasize that during the imitation of an empathetic strategy $E_i$ ($i=0, \ldots, N-1$) a new random number $r$ is generated for the corresponding value of $p_y=q_y$ in order to take into account the role of mutation, which was also considered in the continuous ultimatum game \cite{nowak_s00}. The possibility of slight modifications of the $(p,q)$ values corresponding to a given strategy reduces the sensitivity to initial conditions. Conversely, for the adoption of the strategy $A$ the inherited $(p,q)$ pair remains unchanged. This allows us to study the evolutionary response of the empathetic population to an attack of a given strategy $A$.

\begin{figure}[ht]
\begin{center} \includegraphics[width = 8.5cm]{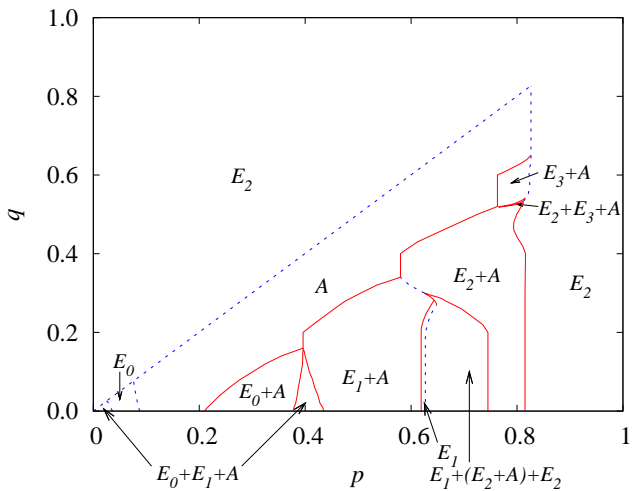}
\caption{\label{phase}Phase diagram for $N=5$ in the presence of strategy $A$ characterized by the parameters $p$ and $q$. Solid red and dashed blue lines denote continuous and discontinuous phase transitions, respectively. Besides several single and two-strategy phases, a rich array of three-strategy phases, in part governed by intricate cyclic dominance, can be observed. Further details are given in the main text.}
\end{center}
\end{figure}

For the evolutionary process the time is measured in Monte Carlo steps (MCS). During one MCS each player has a chance once on average to modify its strategy. For the systematic numerical analysis we have determined the fraction of strategies in the final stationary state when varying the values of $p$ and $q$. For the evaluation of the phase diagram with an adequate accuracy (see Fig.~\ref{phase}), we have used appropriate system sizes, varied from $L=400$ to $6000$, by applying periodic boundary condition. Thermalization and sampling times changed from $t_s,t_{th}=10^4$ to $10^6$ MCS and small variations in the values of $p$ and $q$ are used in the vicinity of phase transitions.

Before presenting the main results, we briefly survey the outcome of the game under well-mixed conditions. In the absence of strategy $A$, the fair $E_2$ strategy prevails over other empathetic strategies \cite{sinatra_jstat09}. The $E_2$ strategy can also dominate strategy $A(p,q)$, but only if $q<0.5$ or $p>0.5$. For $p<0.5$ and $q>0.5$ $E_2$ and $A$ are neutrally stable and subject to random drift, thus always resulting in a fixation if the population is finite.

Figure~\ref{phase} shows the qualitatively distinguishable phases in the final stationary state as a function of $p$ and $q$. If $q>p$ or $p>0.82$ strategy $A$ dies out and the system evolves into a homogeneous state where only the fair $E_2$ strategy remains alive, that is the system reproduces the well-known result reported in previous works \cite{page_bmb02, sanchez_jtb05}, confirming that empathy may lead to the evolution of fairness. Within the complementary triangle, however, there is an intricately webbed region where much more complex outcomes are possible. Besides the single ($A$, $E_0$, $E_1$) and two-strategy [($E_0+A$), ($E_1+A$), ($E_2+A$), ($E_3+A$)] phases, we can observe different three-strategy phases, where the three strategies are caught up in cyclic dominance [($E_0+E_1+A$), ($E_2+E_3+A$)], or even where one strategy in the cycle is an alliance of two strategies [$E_1+(E_{2}+A)+E_2$]. It is emphasized that below the $p=q$ line within the complementary triangle players $A$ invade the whole system if the non-empathetic driving force ($p-q$) is weaker than a threshold value dependent on $p$.

A highlight is also the survivability of strategy $E_3$, which falsifies the assumption that ``superfair'' behavior, when a player offers more than it keeps, is unsustainable. In addition, the fully rational $E_0$ strategy remains alive in three phases (denoted $E_0$, $E_0+A$ and $E_0+E_1+A$), as marked in the corresponding regions of the phase diagram. Apart from this, the intermediate strategy $E_1$ can also dominate the other strategies or coexist with them, either in two- or three-strategy phases. Notice that within the webbed triangle, the fair $E_2$ strategy can never dominate, and it occurs only in rather small-sized regions. From this observation one can conclude that the evolution of fairness by means of empathy is rather vulnerable in the presence of a strategy that is not limited by the $p=q$ condition (in our case strategy $A$). Yet the vulnerability of fairness doesn't stem so much from a direct threat, in the sense that strategy $A$ would be directly superior to strategy $E_2$, but rather from the fact that it elevates the survivability of the other strategies. A direct consequence is a high spatiotemporal complexity of solutions, ranging from simple two-strategy alliances to cyclic dominance between three strategies, as well as cyclic dominance between two strategies and a two-strategy alliance [$E_1+(E_{2}+A)+E_2$].

\begin{figure}[ht]
\begin{center}
\includegraphics[width = 2.83cm]{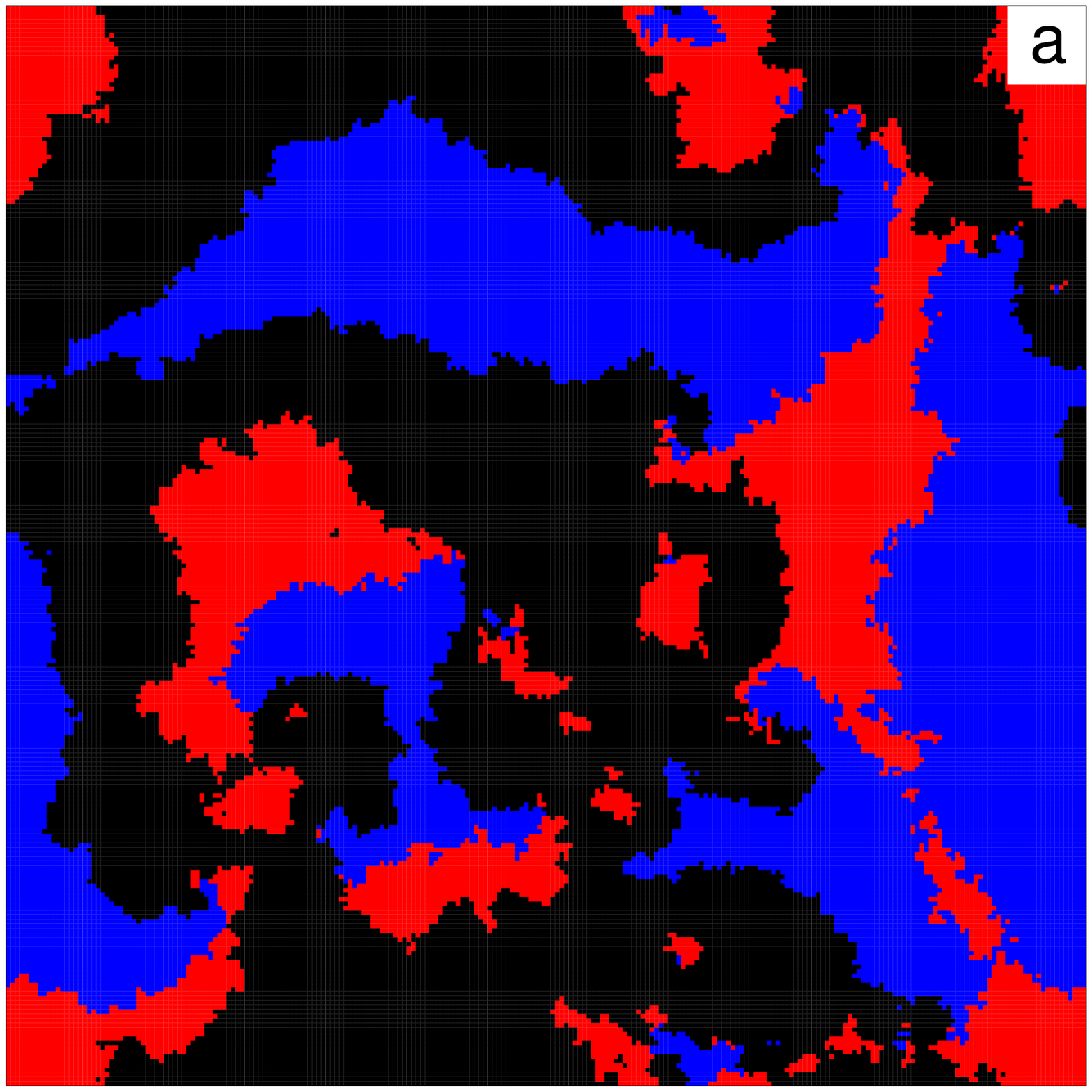}
\includegraphics[width = 2.83cm]{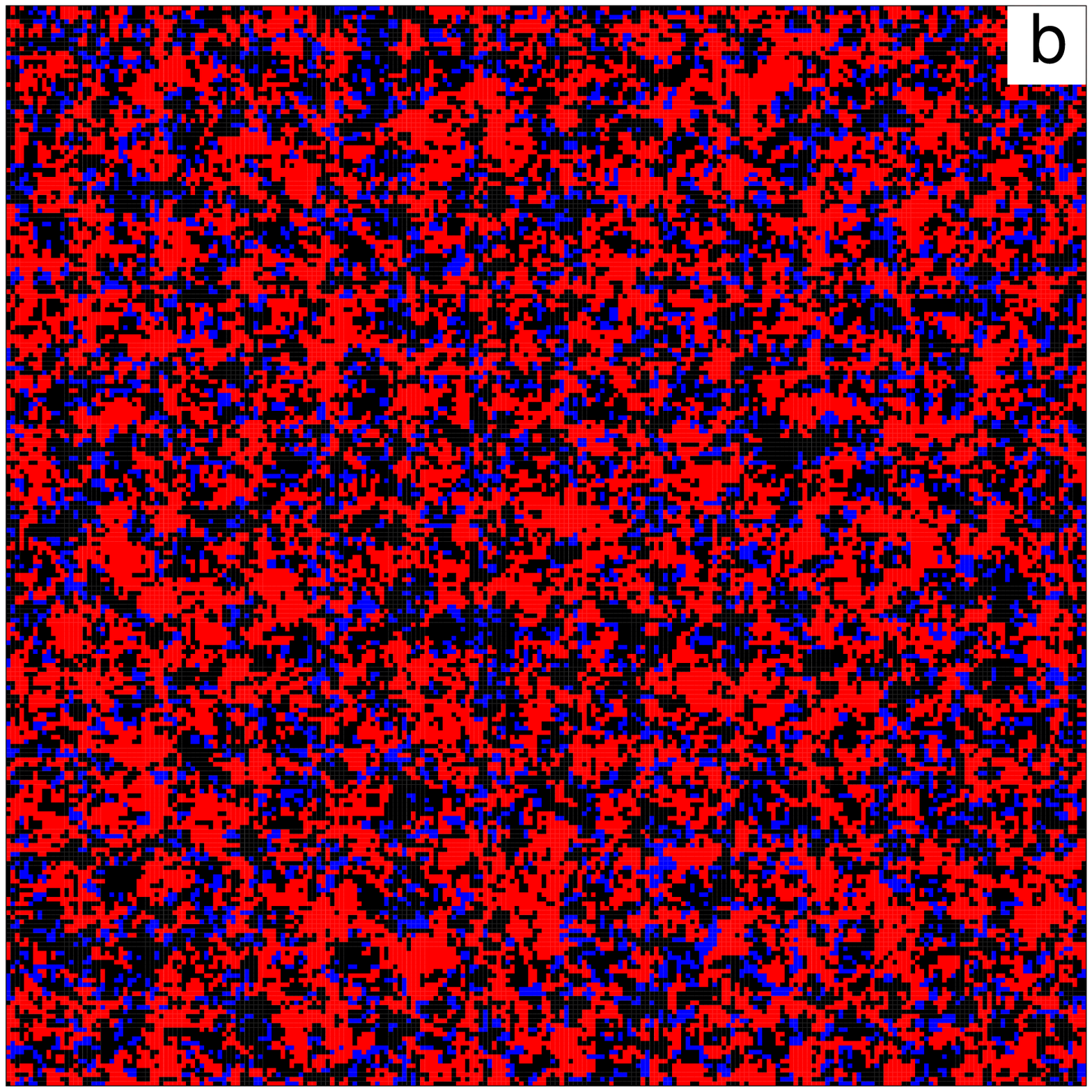}
\includegraphics[width = 2.83cm]{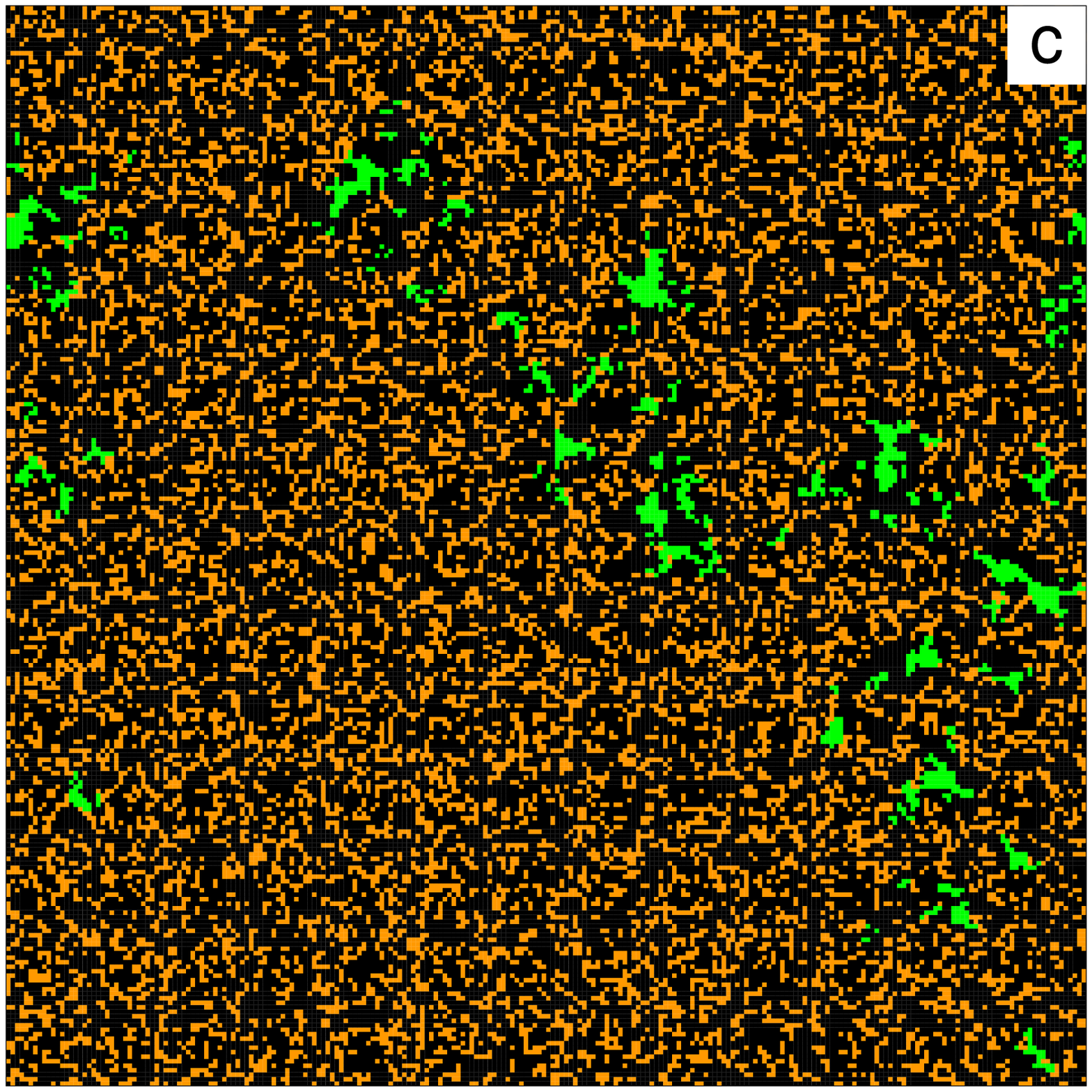}
\caption{\label{snaps}Characteristic spatial patterns emerging for different combinations of $p$ and $q$: (a) $p=0.02$ and $q=0$; (b) $p=0.4$ and $q=0$; (c) $p=0.81$ and $q=0.536$. In all panels the system size is $L=240$ and the strategy colors are those introduced in Fig.~\ref{scheme}.}
\end{center}
\end{figure}

The strategy fractions do not define clearly all the relevant features of the final stationary states, as it is demonstrated by the three snapshots presented in Fig.~\ref{snaps}. Panel (a) of Fig.~\ref{snaps} shows a characteristic snapshot of a self-organizing pattern with traveling invasion fronts and rotating spirals that is related to the spontaneous emergence of cyclic dominance \cite{szabo_pre99} between the three strategies ($E_1 \to E_0 \to A \to E_1$) at $p=0.02$ and $q=0$. The coexistence of the same three strategies is maintained by a fundamentally different mechanism yielding a pattern illustrated in the panel (b) where $p=0.4$ and $q=0$. The latter strategy distribution can be interpreted as a poly-domain structure of the $E_0+A$ and $E_1+A$ phases resembling the phase segregation of a water-oil mixture in the presence of a surfactant \cite{gompper_94}. The latter analogy is supported by the continuous variation (from 0 to 1) in the ratio of the territory of $E_0+A$ and $E_1+A$ phases.

The snapshot (c) in Fig.~\ref{snaps} illustrates another three-strategy state, where strategy $A$ (black) with $p=0.81$ and $q=0.536$ enables the survival of the ``superfair'' $E_3$ strategy (orange), which on the other hand is inferior to $E_2$ (green). The plotted strategy distribution is formed after a long transient process. Namely, within a short time the random strategy distribution evolves into a state dominated by $E_2$ with small growing islands of $A$ strategies. In a small portion of growing $A$ islands several $E_3$ strategies remain alive and their offspring spread away in the whole system when the condition of percolation is satisfied.

\begin{figure}[ht]
\begin{center}
\includegraphics[width = 2.83cm]{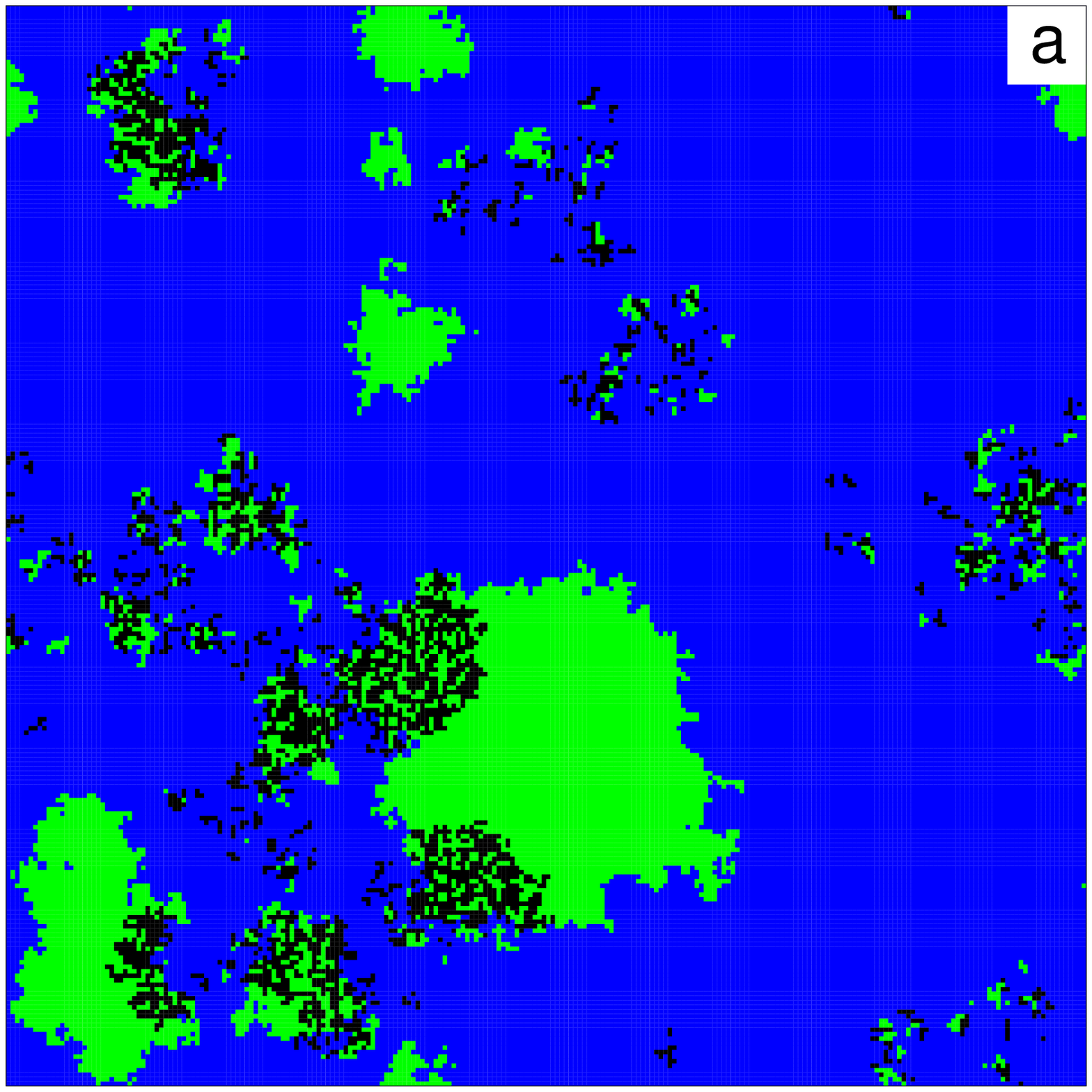}
\includegraphics[width = 2.83cm]{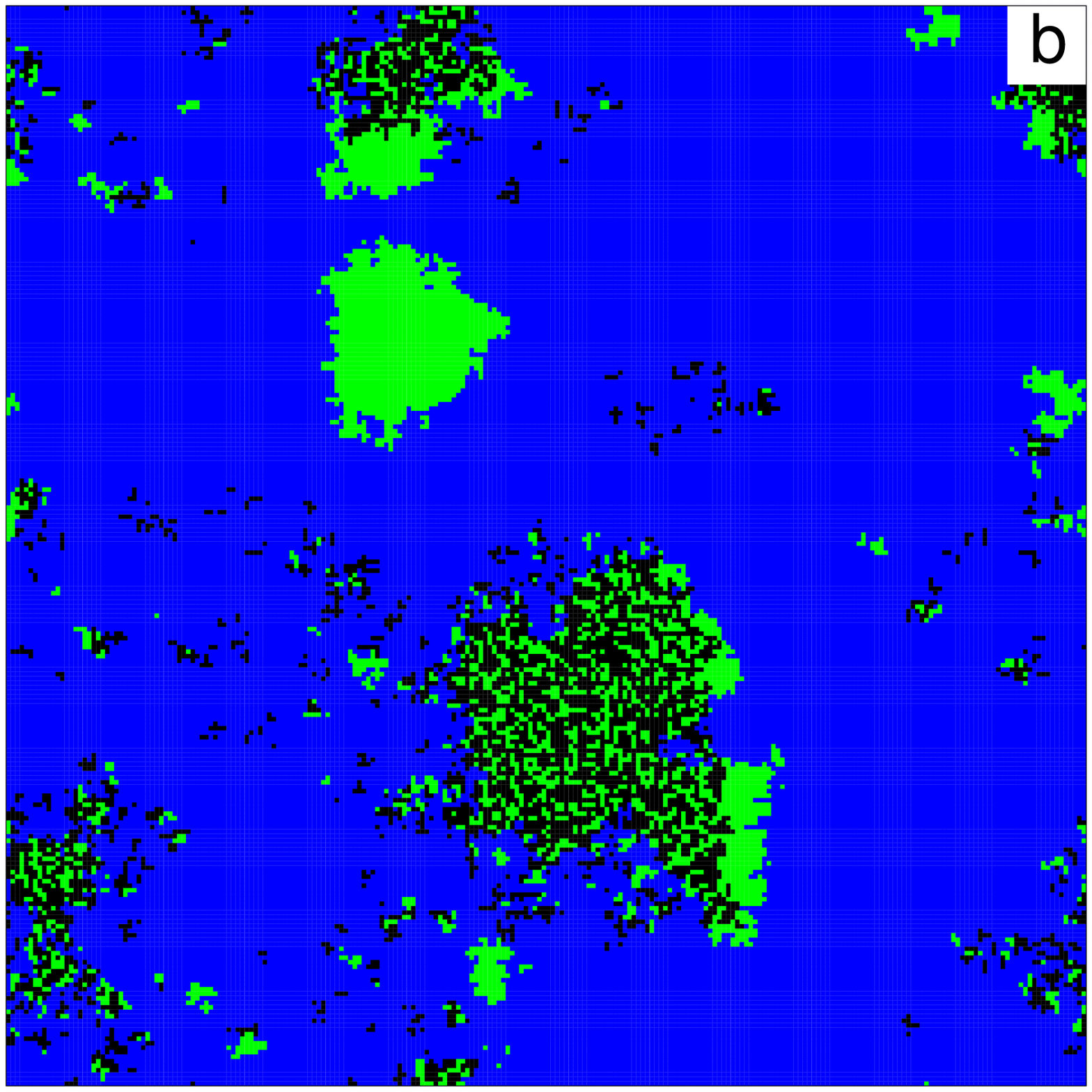}
\includegraphics[width = 2.83cm]{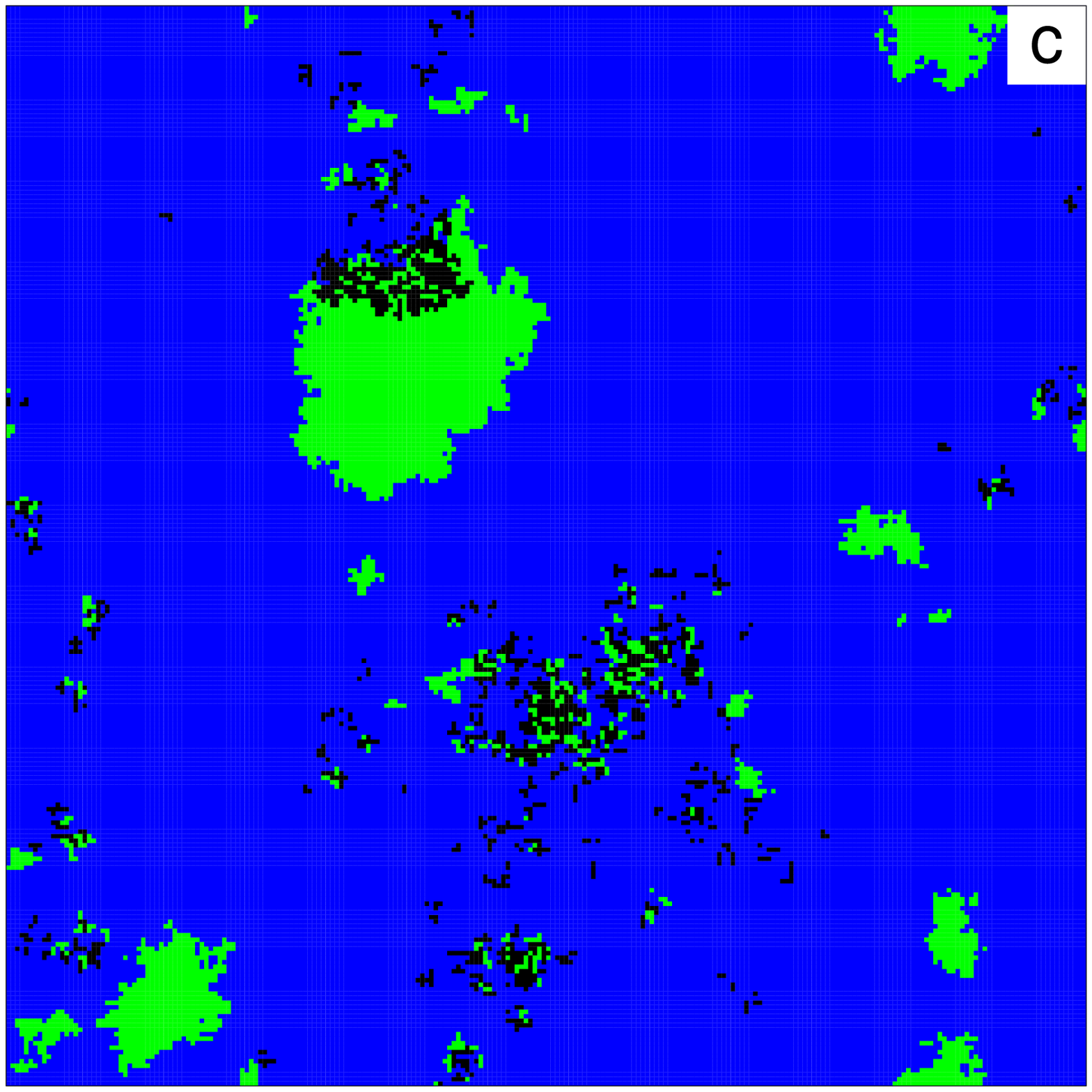}
\caption{\label{alliance}Evolution of the strategy distribution for the exotic $E_1 + (E_{2}+A) + E_2$ phase occurring at $(p,q)=(0.645,0)$. The snapshots were taken at times (a) $t=1000$,(b) $1100$ and (c) $1200$ MCS for $L=240$. The colors of strategies are those introduced in Fig.~\ref{scheme}.}
\end{center}
\end{figure}

The time evolution of another remarkable self-organizing pattern is illustrated by three consecutive snapshots in Fig.~\ref{alliance}, which correspond to the stationary state.
Due to the cyclic dominance ($E_1 \to (E_{2}+A) \to E_2 \to E_1$) emerging for the given $(p,q)$ pair, the islands of $E_2$ players are growing in the sea of $E_1$ players. The growth is blocked once these islands become infected by $A$, which yields the appearance of the $E_2+A$ phase, which in turn can be invaded by $E_1$. In some cases, however, several $E_2$ players survive by forming a sufficiently large nucleon, which closes the loop of cyclic dominance. We note again that in this case the loop of dominance is not formed by three strategies, as for example by the $E_0+E_1+A$ phase [see Fig.~\ref{snaps}(a)], but rather it consists of two strategies ($E_1$ and $E_2$) and an alliance of two strategies ($E_{2}+A$). Although similarly complex phases have been reported before in spatial ecological models \cite{szabo_jtb07} and most recently for the spatial public goods game with pool punishment \cite{szolnoki_pre11}, the current observation enforces that such exotic solutions may be significantly more common than initially assumed, especially in systems describing human behavior.

\begin{figure}[ht]
\begin{center}
\includegraphics[width = 4.25cm]{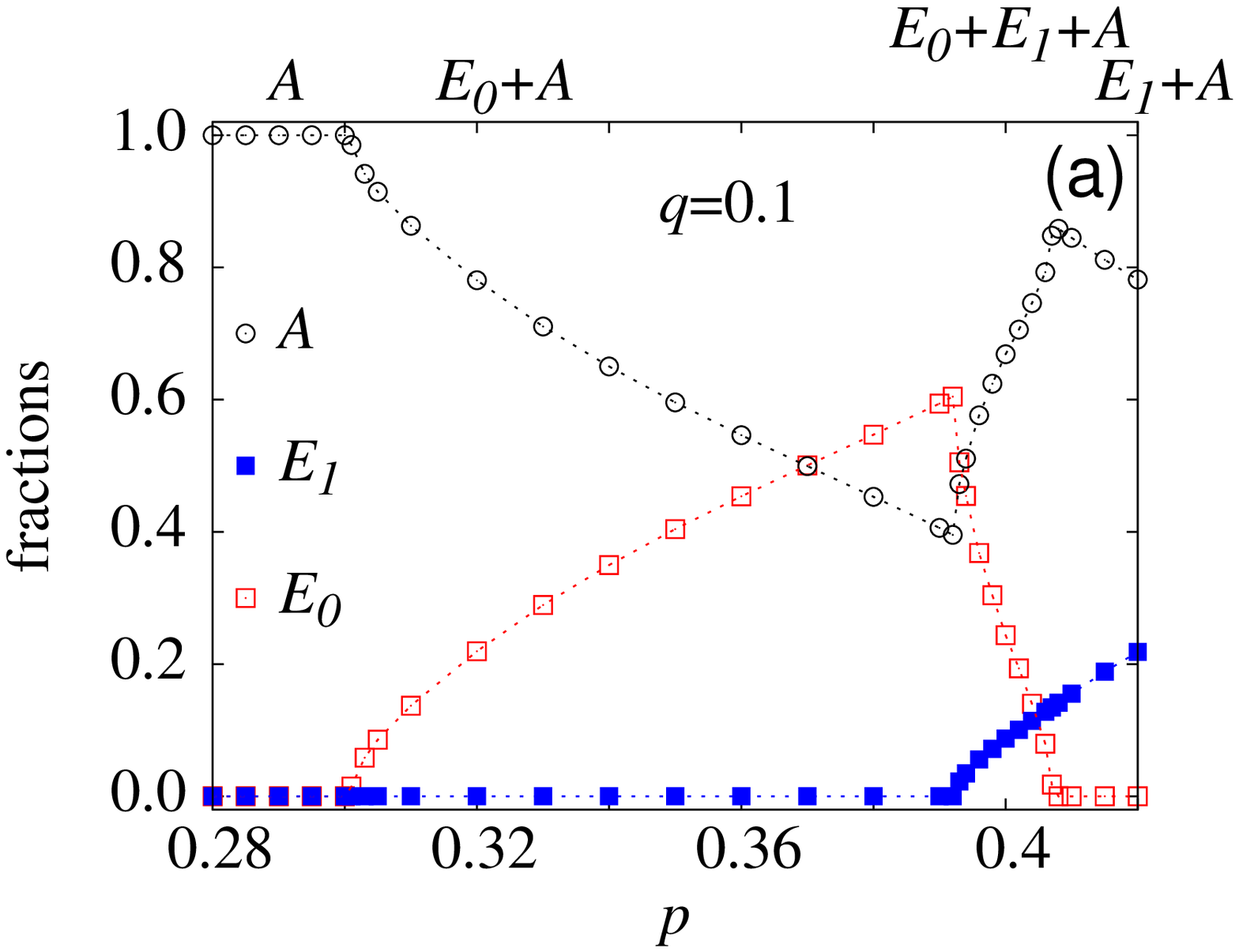}
\includegraphics[width = 4.25cm]{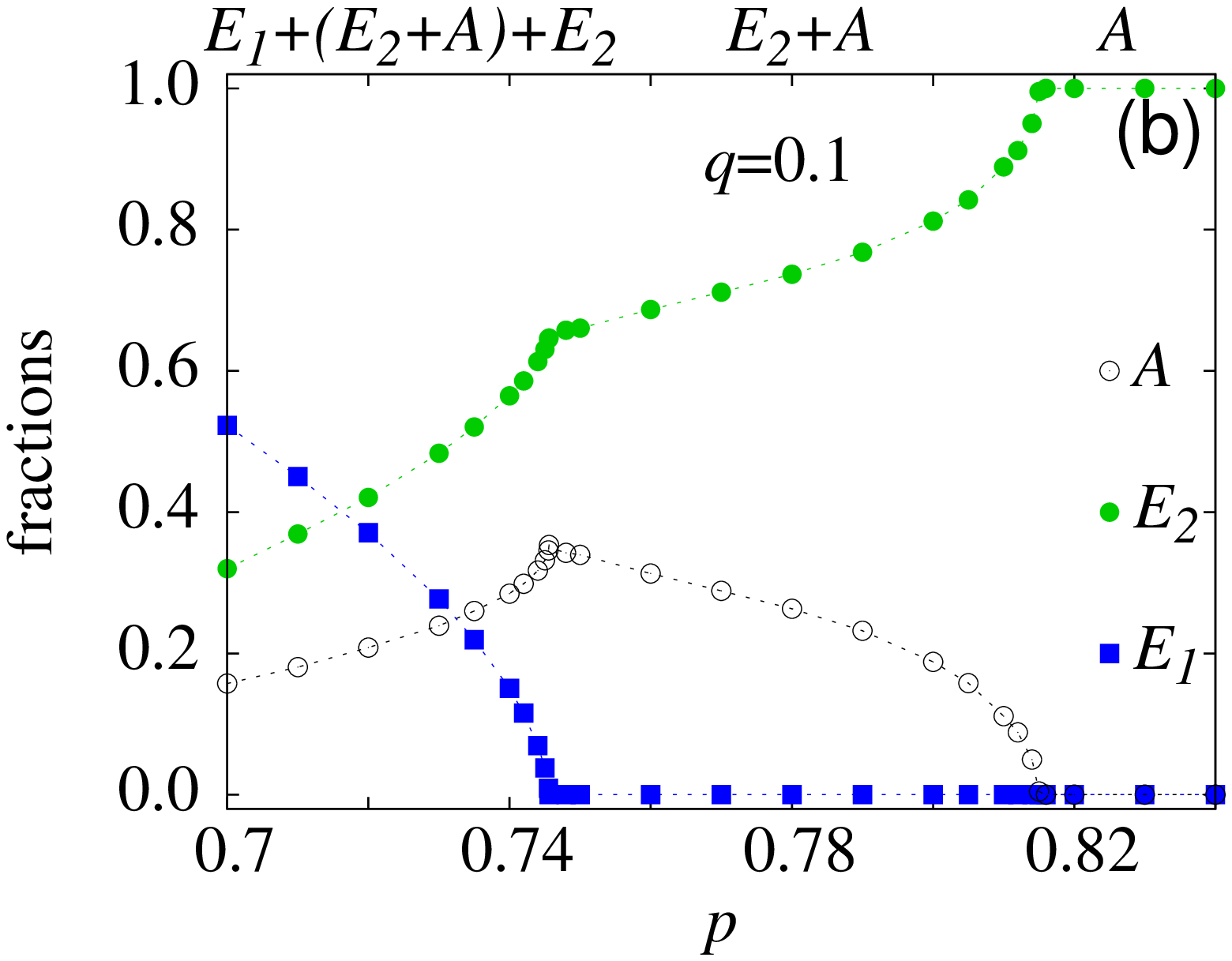}
\includegraphics[width = 4.25cm]{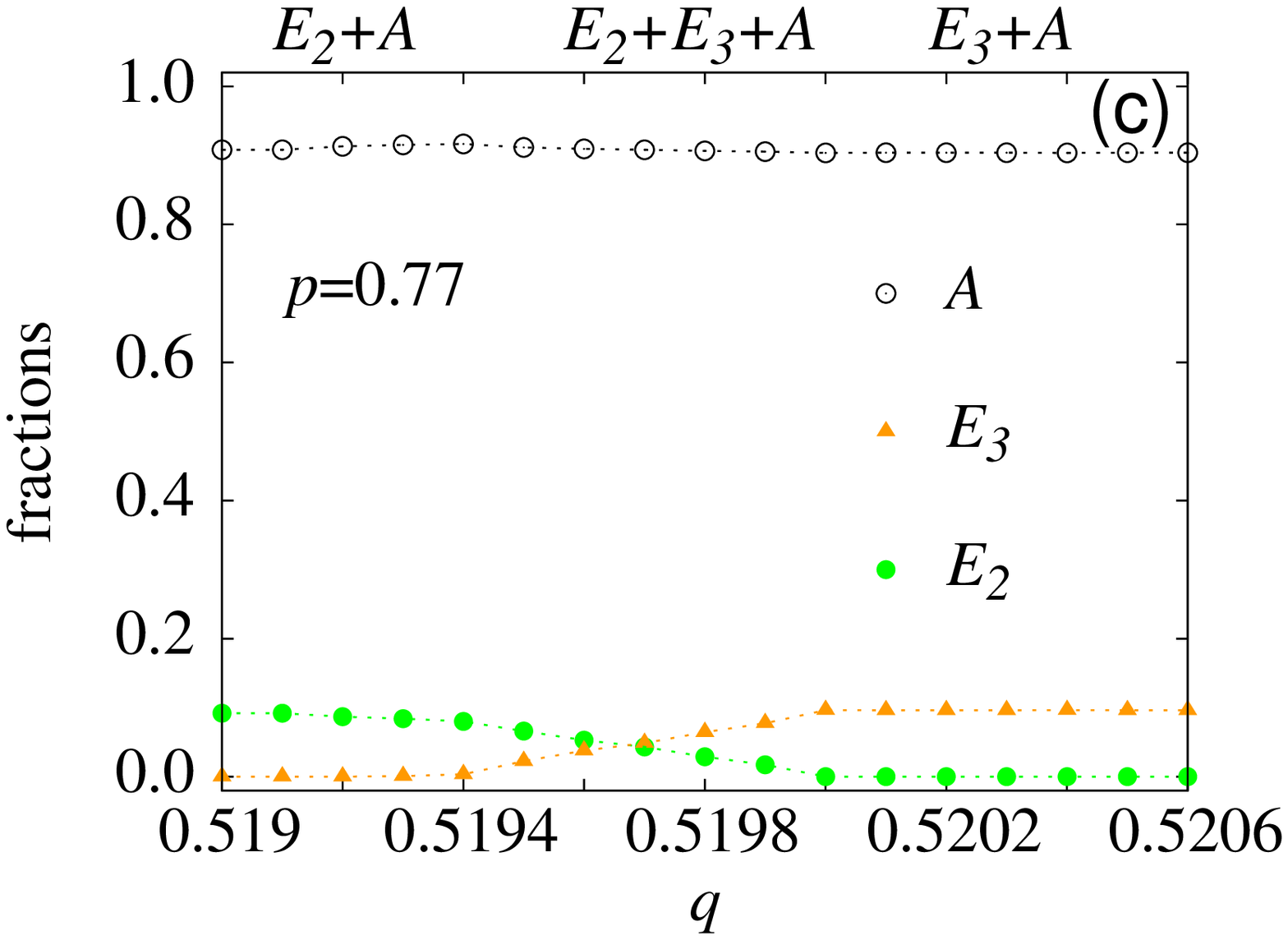}
\includegraphics[width = 4.25cm]{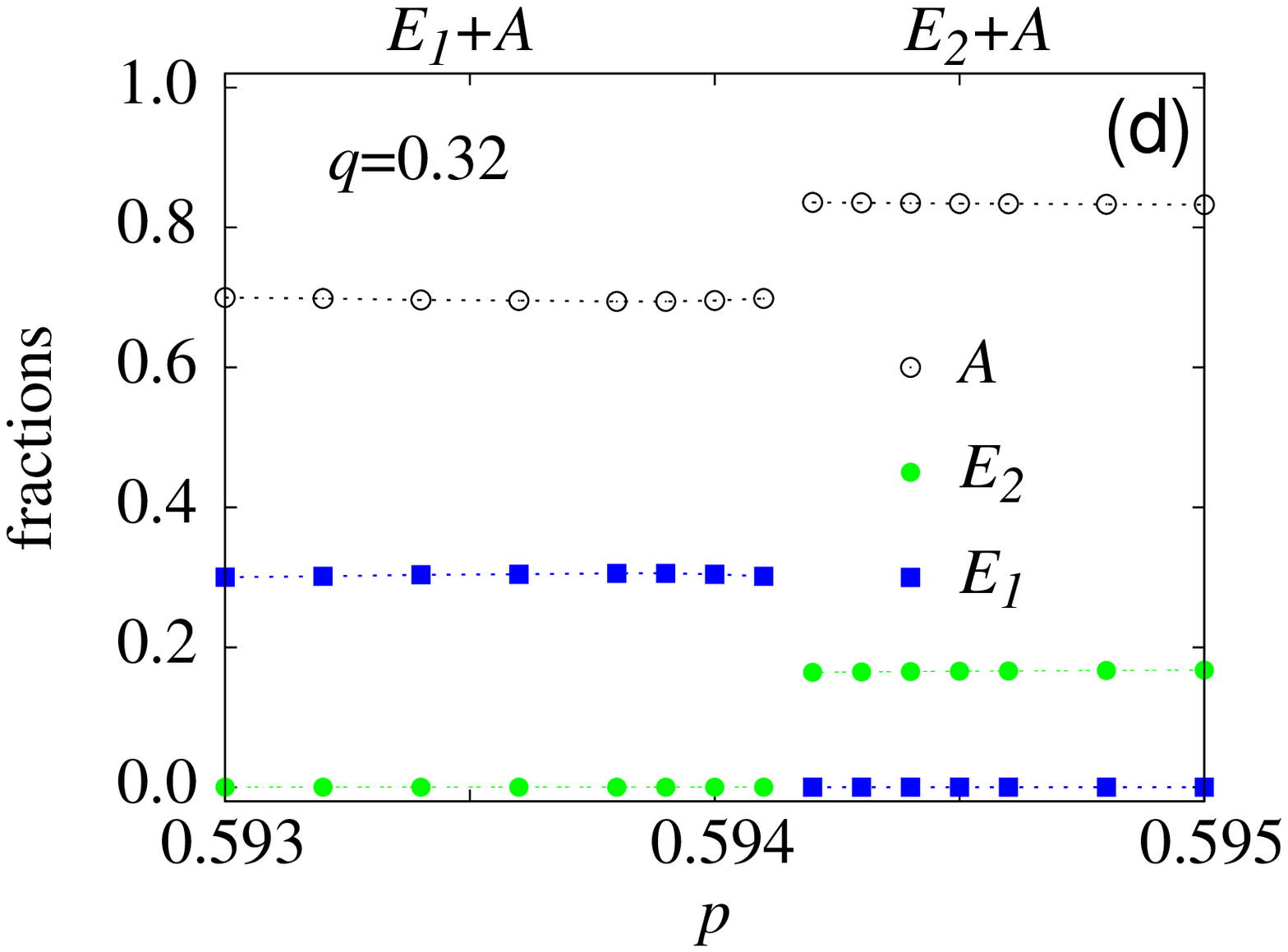}
\includegraphics[width = 4.25cm]{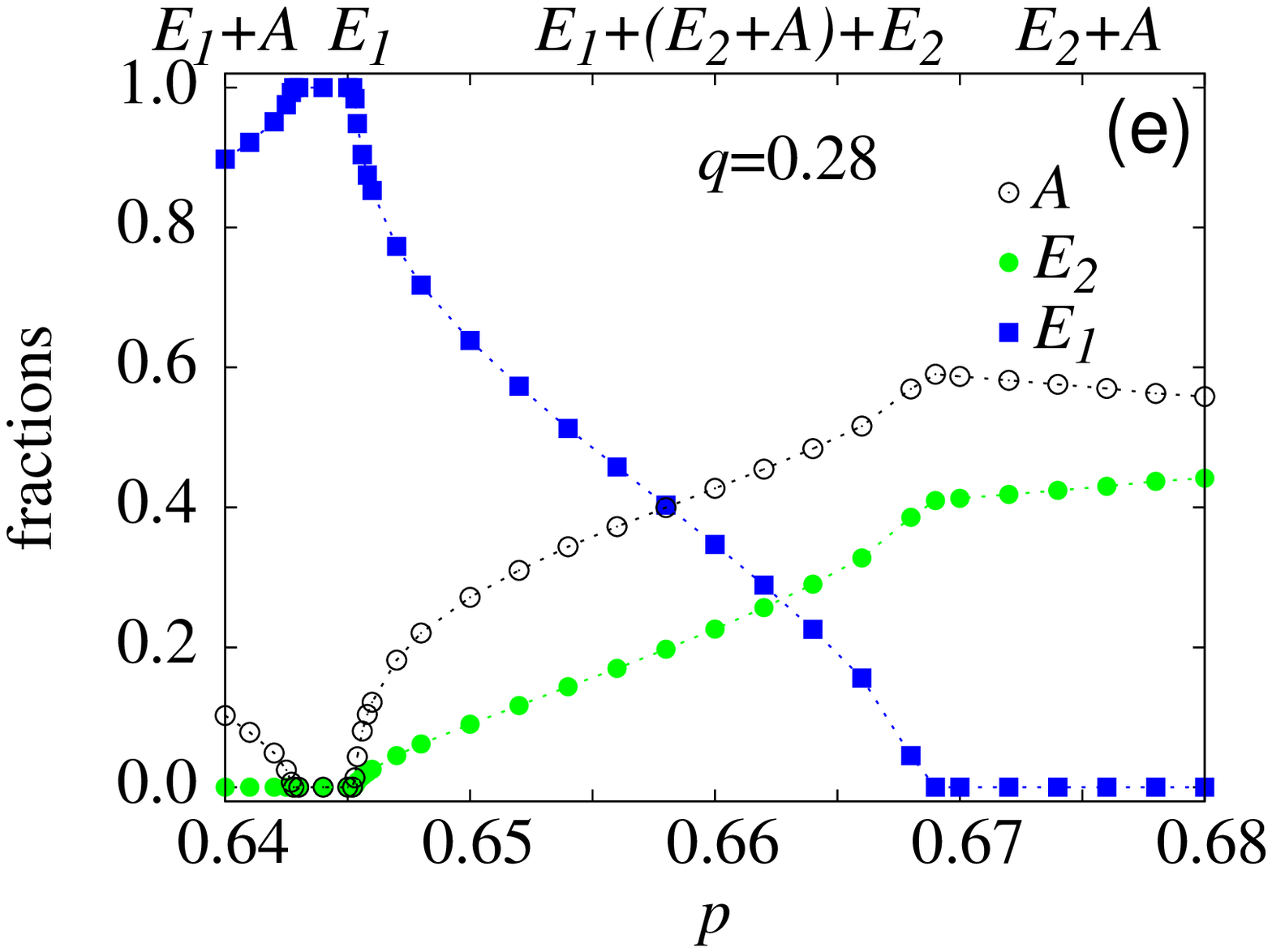}
\includegraphics[width = 4.25cm]{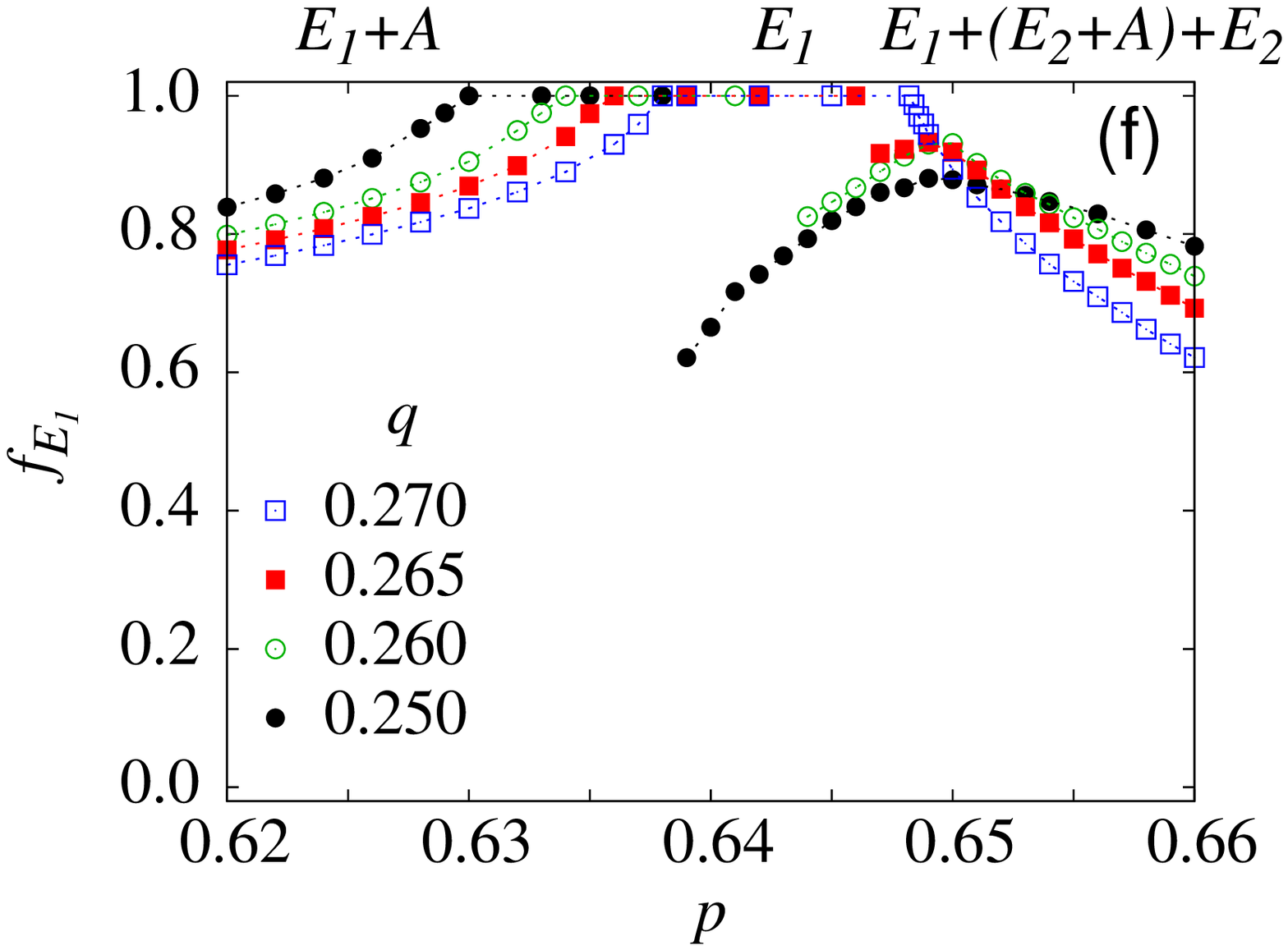}
\caption{\label{cross}Representative cross-sections of the phase diagram depicted in Fig.~\ref{phase} [panels (a-e)], along with the evolution of the order parameter $f_{E_{1}}$ (the fraction of strategy $E_1$) for different values of $q$ [panel (f)]. The symbols indicate results for the strategy fractions as a function of $p$ (or $q$) for fixed $q$ (or $p$) values, as indicated in the corresponding panels.}
\end{center}
\end{figure}

Finally, we examine quantitatively the properties of phase transitions depicted in Fig.~\ref{phase}, which delineate all the different phases that we have graphically described in Figs.~\ref{snaps} and \ref{alliance}. Figure~\ref{cross}(a) shows consecutive continuous phase transitions leading from the pure $A$ phase, over the two-strategy $E_{0}+A$ and the three-strategy $E_{0}+E_{1}+A$ phase, to the two-strategy $E_{1}+A$ phase when $p$ is increased for $q=0.1$. As mentioned above, the three-strategy $E_{2}+E_{3}+A$ phase can be considered as the poly-domain combination of the $E_{0}+A$ and the $E_{1}+A$ phase. Qualitatively similar continuous phase transitions can be observed for larger $p$ values (if $q=0.1$), as illustrated in Fig.~\ref{cross}(b). The two-strategy $E_{2}+A$ and $E_{3}+A$ phases are separated by the three-strategy $E_{2}+E_{2}+A$ phase via continuous transitions if $q$ is increased for $p=0.77$, as shown in Fig.~\ref{cross}(c). On the contrary, Fig.~\ref{cross}(d) represents a discontinuous phase transition from $E_{1}+A$ to $E_{2}+A$. Even more interestingly, the transition line separating the pure $E_1$ phase and the exotic $E_1+(E_{2}+A)+E_2$ phase changes from continuous to discontinuous via a tricritical point at $p=0.6483$ and $q=0.27$. For higher $q$ values the transition is continuous as illustrated in Figure~\ref{cross}(e). By decreasing $q$, however, the transition becomes discontinuous and the jump in the order parameter $f_{E_{1}}$ grows gradually as shown in Fig.~\ref{cross}(f).

In summary, we have proposed and studied a spatial ultimatum game with a discrete set of strategies for a dynamics based on stochastic imitation of a neighboring strategy. The strategy set included $N$ empathetic strategies and an additional $A$ strategy characterized by fixed proposal ($p$) and acceptance ($q$) values. We would like to emphasize that discreteness of possible parameters characterizing empathetic strategies is a natural assumption since human bargain is indeed practically conducted in this way. Our numerical analysis was performed for $N=5$ at a low noise level in order to study the reaction of the empathetic players to the presence of strategy $A$ on the structured population. All the main findings, however, summarized by the phase diagram in Fig.~\ref{phase}, are robust and remain valid also for larger $N$. Naturally, more empathetic strategies offer more coexistence possibilities with the strategy $A$ below the $q=p$ diagonal. We have also tested and confirmed the emergence of alliances for the so-called death-birth strategy updating, where a player is chosen randomly to die and the neighbors compete for the empty site proportional to their payoffs \cite{ohtsuki_jtb06}. This modification too does not alter the possible solutions, and indeed results in a very similar phase diagram. We have thus shown that the previously reported evolution of fairness by means of empathy and spatiality is feasible in the absence of strategy $A$, even for the discrete set of strategies used. Furthermore, we have shown that the introduction of the additional strategy $A$ does not influence the final stationary state for sufficiently large values of $p$ or $q>p$ because players adopting strategy $A$ become extinct within a short transient time, and finally the fair players come to dominate the system. On the contrary, below the $p=q$ line on the $p-q$ plane players $A$ can survive, either alone or by forming complex alliances with the empathetic strategies. Besides the mentioned homogeneous states, we have observed several types of strategy coexistence that were maintained by different mechanisms, primarily routed in the formation of intricate patterns on the spatial grid. Each type of these solutions represents a particular way by means of which a portion of the empathetic players can survive in the presence of non-empathetic behavior. In some cases a non-empathetic strategy can stabilize rational strategies that would be employed by \textit{Homo economicus}, as well as completely irrational strategies that could be adopted by an extremely generous \textit{Homo emoticus}. At the same time, the wide variety of solutions highlights the sensitivity of the system to the parameters of the strategy $A$. The present model supports the spontaneous emergence of strategy associations (composed from a subset of strategies with a specific spatiotemporal structure), which play a fundamental role in many human and biological systems, and as demonstrated in this Letter, can be studied in the theoretical framework of evolutionary games with methods of statistical physics.

\begin{acknowledgments}
This research was supported by the Hungarian National Research Fund (grant K-101490) and the Slovenian Research Agency (grant J1-4055).
\end{acknowledgments}

\end{document}